# Investigating the Impact of Forgetting in Software Development

Utku Ünal
Information Systems
METU
Ankara, Turkey
utku.unal@metu.edu.tr

Eray Tüzün
Computer Engineering
Bilkent University
Ankara, Turkey
eraytuzun@cs.bilkent.edu.tr

Tamer Gezici
Neuroscience
Bilkent University
Ankara, Turkey
tamergezici@gmail.com

Ausaf Ahmed Farooqui
Psychology
Bilkent University
Ankara, Turkey
ausaf.farooqui@gmail.com

## ABSTRACT

*Context:* Forgetting is defined as a gradual process of losing information. Even though there are many studies demonstrating the effect of forgetting in software development, to the best of our knowledge, no study explores the impact of forgetting in software development using a controlled experiment approach.

*Objective:* We would like to provide insights on the impact of forgetting in software development projects. We want to examine whether the recency & frequency of interaction impact forgetting in software development.

*Methods:* We will conduct an experiment that examines the impact of forgetting in software development. Participants will first do an initial task. According to their initial task performance, they will be assigned to either the experiment or the control group. The experiment group will then do two additional tasks to enhance their exposure to the code. Both groups will then do a final task to see if additional exposure to the code benefits the experiment group's performance in the final task. Finally, we will conduct a survey and a recall task with the same participants to collect data about their perceptions of forgetting and quantify their memory performance, respectively.

## 1  INTRODUCTION

The elapsed time to fix bugs and to develop new features is vital to the software development process. Developers that are most familiar with the code should be assigned to have quicker and better completion of tasks. Forgetting the source code over time is an important factor that influences code familiarity of developers. Ebbinghaus [1] is the pioneer of forgetting studies, and his ideas have been widely accepted and successfully replicated [2]. He also was the first to suggest a measurement of forgetting, called the forgetting curve, which shows how information is lost over time if there is no extra effort to retain it. Starting with Wright [3], the forgetting curve's counterpart, the learning curve, has been observed in different industrial processes and its effects have improved estimation of production.

In the software engineering community, the forgetting and learning curves were explored much later [4, 5]. Several studies mined software repositories [6, 7, 8] to measure the effects of forgetting and learning curves in software development. The main problem with the mining studies is the fact that there are many potential confounding variables that the researchers could not control, which may interfere with their conclusions. There has also been a recent study surveying the developers of software projects on file familiarity [9]. The authors assert that the forgetting curve of Ebbinghaus [1] is also applicable in software development and there is a comparatively strong relationship between repetitions and familiarity of source codes. Although these results are promising, the conclusions can be limited because the underlying data was gathered via survey, so it only indicates the perception of code authors about forgetting.

In software development, assigning individual developers to tasks is a common scenario. One of the main objectives of making such assignments should be to minimize the effort required to complete each task [10]. If we can verify that repetitive and recent exposure to the codebase helps develop faster and better-quality code, team leaders can utilize this finding to assign developers best suited to given tasks. In this study, to complement and compare the findings of earlier studies, we propose a controlled experiment where an experiment group and a control group get exposed to the source code in different frequencies and time intervals to assess the impact of forgetting in software development.

In Section 2, we discuss our research questions and the corresponding hypotheses. In Section 3, we describe the research protocol. In Section 4, we present the results of the pilot study that we conducted, and in Section 5, we discuss the threats to validity. Finally, in Section 6, we compare related work with our study.

## 2 RESEARCH QUESTIONS AND HYPOTHESES

The process of interacting with files when solving bugs and developing new features is central to software development. Our dependent variable is how much time the developer spends on solving a task (time to resolution (TTR)), considering previous interaction (s) with the same file.

Previous studies on forgetting have observed a direct relation between time-since-learning and retention of information [1, 2]. In software development studies, Krüger et al. [9] also found through a survey that the more time elapses since the last interaction with a file, the less confident the developers are about its contents. Hence, in this study, we explore the following research question:

Table 1. The defined variables of the study.

| Name | Description | Operationalization | Task Type |
| --- | --- | --- | --- |
| **Time of Inactivity (TI)** | The time since the participant's last interaction with the code. | Measuring the time elapsed since the participant's last pull request. | Independent Variable |
| **Number of Exposure (NE)** | How many times participants were exposed to the code? | Counting the tasks or bugs assigned to the participant. | Independent Variable |
| **Time to resolve (TTR)** | The time participants spend solving the task or bug assigned to them. | Measuring the time from when a participant changes their task status to "In progress" until the participant creates the pull request for the related task. | Dependent Variable |
| **Initial Task Performance (ITP)** | Participant's initial task performance | Measuring the completion times for the initial task. | Independent Variable |

**RQ1:** *What is the relationship between the developer's interaction with the source code and forgetting that source code?*

**H1:** *The more frequent a developer's interaction with a source code is, the less likely is the developer to forget that source code.*

Psychologists have found that actively [11] or attentively [12] interacting with information improves the likelihood that that information will not be forgotten.

## 3 RESEARCH PROTOCOL

### 3.1 Variables & Notations

The variables used in our experiment are presented in Table 1. The control and experiment groups receive different numbers of tasks (an activity that would require coding) with different frequencies. This creates different numbers of exposure (NE) to the code as well different times of inactivity (TI) since the last exposure. NE and TI are our independent variables for RQ1. Time to resolve a subsequent coding task (TTR) is our vital dependent variable since it indicates the participants' performance in solving coding related problems. We, therefore, use TTR to monitor the participants' performance in tasks throughout the experiment.

### 3.2 Materials & Objects

Before we start the experiment, we will conduct a survey to get relevant participant background information. We will use Google Forms for this. The experiment contains improvement and bug-fixing tasks. Participants will be assigned to relevant tasks and are expected to complete them. The project will be in Java. To track tasks and bugs in our experiment, we will create a project board on GitHub, and will use it as an issue-tracking system. When the issue is assigned to a participant, we expect them to create a new branch and commit to it. We use Git as a version control system and GitHub as a service, which is used to manage Git repositories for these types of tasks. Furthermore, we will also prepare a wiki page in GitHub with instructions for setting up the development environment. This page will also have a frequently asked questions (FAQ) section. The participants will be able to create new branches for new issues and commit their changes to them; they can then create pull requests for code reviews. The participants should have prior experience with Java, and they should be familiar with basic Git commands.

In the pre-survey, there will be questions regarding participants' familiarities with Java and Git. Since our participants are senior year computer science students at the same university, we expect their knowledge levels to be very close to each other. Although we do not expect major differences across participants, we will assign them to different groups according to their initial task performance (ITP).

The assigned tasks will take an average of 2 hours to complete (according to the pilot study results). It also takes an extra 10 or 15 minutes on average for each participant to pull the code from GitHub, create a new branch, and push after a commit. This means that if we do not include surveys, control group participants would allocate 4 hours and 30 minutes to the experiment as they only take the first and final tasks. Since the experiment group participants have two more tasks than the control group participants, this period is expected to be around 9 hours. To motivate people to participate in the experiment, we will hold a prize draw among participants. At the end of our experiment, we will present the Amazon Gift Cards to 5 participants who won the lottery. To further motivate the participants and reduce their attrition rate, we will register the experiment with Bilkent University and ensure that the students get credits toward a GE 250 / 251 Collegiate Activities class. This study is approved by the Middle East Technical University Ethics Committee (Approval No: 092-ODTU-2021).

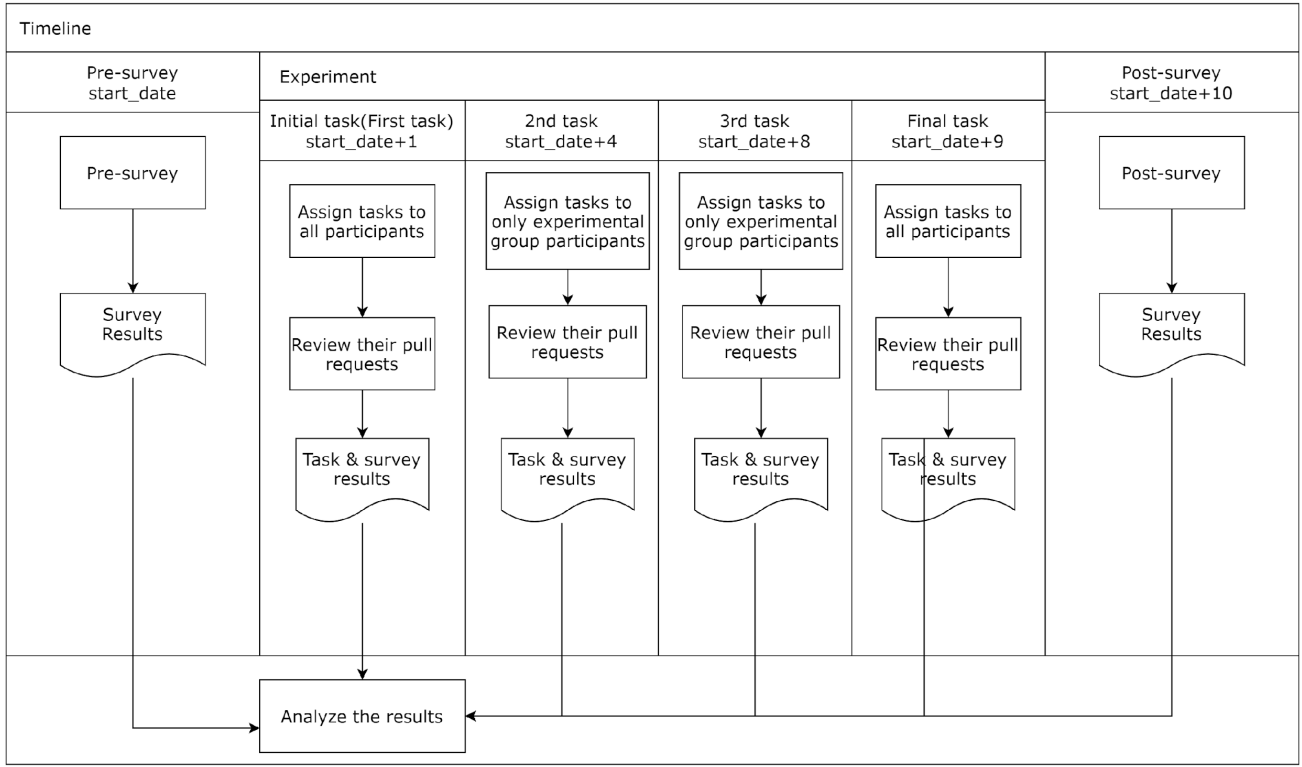

**Fig.1. Execution plan diagram of the experiment**

## 3.2    Participants

We plan to perform our experiment with 40 participants (20 each in control and experiment groups). Our participants will consist of senior computer science students from Bilkent University. As a prerequisite they have to be familiar with Java and Git.

## 3.3    Execution Plan

Two of the key parameters in the experiment are the overall experiment duration and the duration of intervals for consecutive tasks. While determining the duration of the intervals, our aim is not to decrease our experiment group's retention rate too much and to expose them to the code by giving them more tasks.

We benefited from Karpicke & Roediger [13] on this issue; they found that repeated testing produced superior long-term retention after a two-day delay. Therefore, in the experiment, we will give the tasks to our participants at least two-day intervals. When the participants complete their tasks, their access to the project will be suspended until the next task; we do not want them to access the project and examine the code except for the period we allow.

As shown in Figure 1, after the pre-survey part, we start our experiment with the initial task. Then, according to the initial task results, we assign our participants into two groups:

- *Control Group:* The members of this group will be assigned to only the initial and the final tasks during the experiment. Since there are eight days between the initial and the final task, they will not interact with the code for eight days.

- *Experiment Group:* The members of this group will be assigned four tasks during the experiment. They will be assigned two more tasks than the control group during the experiment.

After grouping the participants, the experiment begins; we expect the participants to complete three necessary steps per task during the experiment. These are creating new branches for each issue, changing the task's status as "In Progress" and creating the pull requests after completing the task. After each task, the participant's work is evaluated, and its results are reported. The evaluation includes running unit tests for correctness.

We will start with our initial task which requires participants to comprehensively visit all components of the project. Since the participants will have to trace the whole project, they will gain initial familiarity with all its components. After the first task, we will assign the two additional tasks to the experiment group according to the experiment timeline. The final task assigned to both groups will again be comprehensive like the initial task, but it will be more difficult than the initial task in order to be sensitive to differences in participants' abilities.

The tasks here do not require high algorithmic knowledge. Instead, they require many transitions between files and require making several changes in each file. This would increase the familiarity of the participants with the codebase as well as test their familiarity in the final task. An example task can be seen in Figure 2.

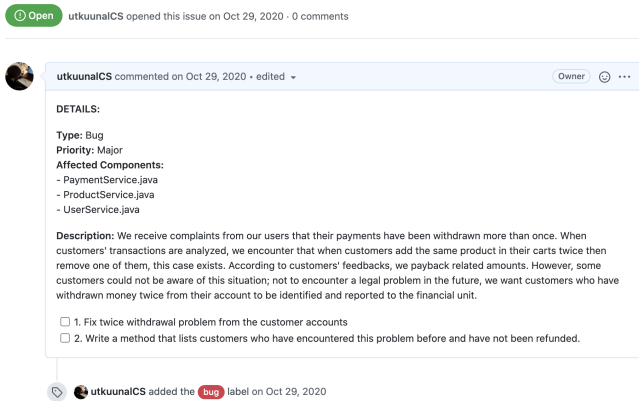

**Fig. 2. Example task format in GitHub**

Before participants start the assigned tasks to them, they should create new branches for tasks from their development branch and make initial commits by labeling their issues' IDs. Thus, the status of the task assigned to them will automatically change to "In Progress". When an issue moves to "In Progress", it is regarded as the time a participant officially started the issue. This change can be seen in Figure 3. Furthermore, we provide different branches for each participant in the experiment and observe their progress individually. We also restrict participants' access to others' branches.

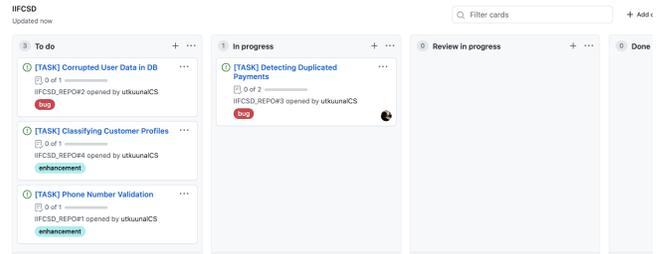

**Fig. 3. Status change when starting the task**

The timestamp of participants' last commits is considered as the completion time of the task as shown in Figure 4. When the participants make their last commit to the issue, they create pull requests for the code review as shown in Figure 5.

After the experiment process is completed, post-survey starts. The post-survey consists of three categories of questions:

● Questions that assess the difficulty, familiarity, and comprehensiveness of the task. (These questions were asked to the pilot study groups to calibrate the questions)

● Questions related to the perception of participants about the forgetting effect.

● Questions related to how well the participants remember specific parts of the code.

Since we will perform the post-survey after the experiment, we do not foresee any kind of bias created because of the survey.

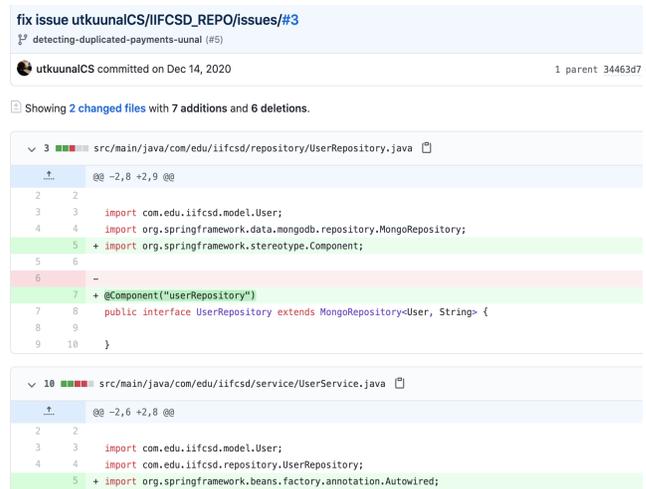

**Fig. 4. Final commit to represent the task is completed**

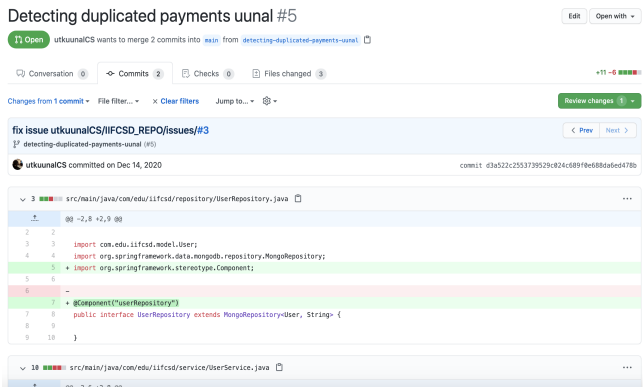

**Fig. 5. Creating a pull request for code review**

## 3.4    Evaluation of the Tasks

We will be enforcing task correctness throughout the experiment for both control and experiment groups. When a participant submits their pull-request (PR), we will be running a set of predetermined unit tests to run for their correctness. The participants will be instructed to modify their codes (corresponding PR for that task) if the PR does not pass all the tests. That way we will ensure that we have a working solution per each task.

We will be recording the time per each successful PR acceptance. Since the evaluation of all the tasks are automated via running predetermined unit tests, we do not expect any kind of bias towards the different experiment groups.

## 3.5    Analysis Plan

1.       We will present the descriptive statistics such as means, medians, standard deviations, max, and min values for each dependent and independent variable in our experiment by creating a separate table for each task and group. We will also present the scatter plots to show potential correlations between these examined variables.

2.       We will use inferential statistics to examine the experimental results and validate our hypotheses. First, we will use the Shapiro-Wilks to test for normality. If there is normal distribution, we use the t-test. Otherwise, we use the Mann-Whitney U test as a nonparametric one. After how significant the differences between groups are analyzed, we also calculate effect size with Cohen's d. If the standard deviations are significantly different between groups, we prefer Glass's delta instead of Cohen's d to measure effect size. Our analysis plan is also demonstrated in Figure 6.

Our H1 asserts that the higher NE is / lower TI is, the more likely the participants will remember the code. We expect the experiment group participants who have interacted with the code recently and have been exposed to the code many times, should complete the final task with a lower TTR than the control group participants. We also expect groups to differ in variables such as task completion, and LOC. To verify our H1, we check to see any significant changes between groups depending on the TI and NE.

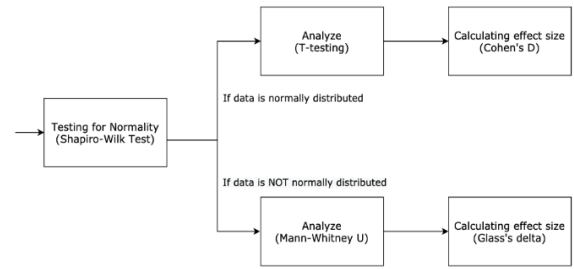

**Fig. 6. Analysis Plan**

## 4   PILOT STUDY

We conducted a pilot study with four participants to fine-tune our design. Participants were computer science graduates with Java experience. The Control and Experiment group were instructed via separate Zoom meetings. Task 1 was given to both groups on day 1. Subsequently, the Experiment group did two additional tasks (Tasks 2 and 3) on days 4 and 7. Finally, both groups did the final Task 4 on day 8. They used the same programming environment. During the task execution, participants could ask questions via text messages, voice chat or email. Each query was noted down. The relevant ones will be added to a FAQ section for the main experiment. A post-survey was conducted the day after the experiment was completed.

We conducted follow-up interviews with participants for feedback. Their comments and suggestions related to the surveys, experiment setup and task descriptions were considered. We summarized the lessons learned from this pilot study as follows:

**1. Environmental Problems**

The project has many installation dependencies which caused participants to struggle setting their working environment during the first task. For this reason, we think that participants' first task performances might have been affected negatively. We encountered versioning problems which are related with Java SDK and Intellij IDE. During our actual experimental work, we plan to hold a training session with all participants prior to the experiment to resolve potential environmental issues and get them to focus immediately on the experiment. As a solution, by using an interactive learning platform, participants will be given a ready-for-development environment in which they can start the task immediately. Thus, participants may elude from related to setting up the development environment.

**2. Incomprehensible Task Descriptions**

Before the experiment, a Wiki section has been prepared in the repositories. The Wiki section contained a general tutorial explaining how to set up their coding environment and a FAQ section. During the experiment, participants were given the opportunity to ask questions regarding the unclear task descriptions for clarification. All questions which were asked by the participants were recorded and will be added to the FAQ section for future use. Also, some task descriptions will be edited to minimize confusion.

### 3. Participant's background

Participant performance also is affected by individual coding skills. Our participant, whose user code is "aBo7hHk5", completed the first task much earlier than other participants. When we look at their survey responses, we see that he has more development experience than other candidates. We noted this and decided to use participants' initial task performance to evenly distribute the participants to the groups depending on their skills.

### 4. Unit Tests

Three of our participants stated that there were insufficient number of unit tests. We noted this and plan to increase the number of unit tests per task.

Table 2 shows the results of this pilot study. As expected, the experiment group with more exposure to the code (because they had done additional tasks 2 and 3) were a lot faster in completing the final task compared to the control group with less exposure to the code.

## 5 THREATS TO VALIDITY

The purpose of our experiment is not to measure knowledge of a particular area or technology. Therefore, task contents will not be based on a specific domain or a specific Java knowledge but rather based on essential algorithm development and debugging knowledge. Thus, we attempt to minimize the effect of initial familiarity, due to previous experience as much as possible. Since we will proceed with a Java-based web project in our experiment, differences may arise due to initial differences of coding skill set. Namely, the participants with previous development experience will have an advantage compared to a participant who has not developed code in this area before. To minimize the variance threat among the participants we decided to select our participants from students with a similar background.

We further minimize the variance threat by distributing the participants to different groups according to their ITP. As a result of this categorization, we aim to create more homogenous experiment groups.

To respond to RQ1, we observe when and how often participants last worked with code in our experiment. We assume that participants have no contact with the code at the time intervals we specified.

**Table 2. Pilot study task performances.**

|  | Initial Task (Task 1) |  | Task 2 |  | Task 3 |  | Final Task (Task 4) |  |
| --- | --- | --- | --- | --- | --- | --- | --- | --- |
| Participant's user code | TTR (minutes) | Unit Tests Passed | TTR (minutes) | Unit Tests Passed | TTR (minutes) | Unit Tests Passed | TTR (minutes) | Unit Tests Passed |
| Experiment Group P1, aBo7hHk5 | 136 | 5/5 | 53 | 3/3 | 42 | 1/1 | 73 | 12/12 |
| Experiment Group P2, L2T84YZ6 | 223 | 5/5 | 146 | 3/3 | 35 | 1/1 | 99 | 12/12 |
| Control Group P1, D7CEEJ6e | 235 | 2/5 | X | X | X | X | 175 | 12/12 |
| Control Group P2, H3eyMgVZ | 299 | 5/5 | X | X | X | X | 229 | 12/12 |

As a precaution to continue with these assumptions, we do not allow access to the source code beyond the time allowed. Their access to the source code is suspended after their tasks are completed. They also do not have access to each other's code. Thus, they cannot interact with the code or examine other people's code or solutions, except for the time given to them.

While we aim to do our experiment physically, we will perform it online due to the pandemic conditions. Therefore, we will not be able to physically observe the participants during each stage of the experiment. Because of the online setting, participants may potentially violate the experiment rules. For example, the participant can solve the task with the assistance of another person. They can also manipulate time taken to complete the task by changing the task's status to "In Progress" after they have already started solving the task. Considering all of these, no matter how many precautions are taken, participants can sidestep any precaution we take at some point if they wish. Therefore, we consider asking our participants to sign an honor agreement at the beginning of our experiment.

Another potential threat is the fact that the time of completing a task includes not only remembering the code (or the architecture of the project), but involves as well knowledge on the process the development tools involved to solve the task (GitHub, Issue Tracking System, etc.). The experiment group that has performed 4 tasks will be better trained on these tools than the control group, those that have just performed 2 tasks, resulting in fulfilling the tasks in less time. However, this gain in time will not only be because of having forgotten less, but partly because of more familiarity with the toolchain. To prevent this possible threat, we will prepare tutorials about the tools which will be used in the experiment. These tutorials contain all possible cases which participants might encounter.

When we list the possible threats to construct validity in our experiment, they are "bias in experimental design", and "hypothesis guessing". For the "bias in experimental design" threat, in our experiment, we do not particularly search for a specific type of participants (i.e. the most skillful coder between the participants). To prevent this possible bias, participants' personal information will be anonymized, and personal information will not be presented when the experiment results are analyzed and shared. Since all the data is anonymous and the analysis will be conducted automatically, we do not foresee the existence of bias.

The other possible threat for the construct validity is "hypothesis guessing" is where the experiment participants try to guess what the real purpose of the study is and may change their reactions accordingly. For this threat, limited information will be given about the experiment. They will only be informed that tasks will be assigned to them, and they will be asked to solve them at certain time intervals through certain tools.

## 6 RELATED WORKS

There are several related works exploring the forgetting curve in different domain contexts. Other complementary works include empirical studies and data analysis based on company projects or open-source software projects.

In the IT Service domain, to investigate the impact of the learning and forgetting curve for different types of knowledge, including domain, technology, and methodology, Kang and Hahn [14] performed a retrospective analysis over the archival company data from a sample of 556 software projects. According to this study's findings, previous experience based on the same methodology or technology has a more significant impact on software project performance than those in the same application area. This study demonstrates that a methodology or technology-based experience can positively reflect the participants' experiment performance. Therefore, we keep the participants' Java familiarity as a different metric in our analysis. We try to keep our tasks at a basic algorithm development level by decreasing methodology and technology effects on our tasks to avoid differentiation in the participants' performances depending on this parameter.

In open-source software projects, to investigate the impact of the learning curve and understand the factors that affect the learning, Au et al. [15] analyzed the number of resolved bugs and the resolution time of 118 different open-source software (OSS) projects. The result of this study showed that small-sized teams learn faster. Additionally, project performance increases accordingly when more bugs are assigned to specific developers. In our study, independent of team sizes, we try to observe the change in participants' performance by assigning more tasks to specific developers.

The other complementary study in the OSS projects domain, to investigate the learning curve effect in open-source software projects, Huntley [6] analyzed the bug life cycle time from Apache and Mozilla projects datasets. One of the critical outcomes expressed from this study is that the success of any given learning approach might be situational and should be analyzed in the context of the related organization. On the other hand, in our study, we set up our experiment independent of organizational criteria. We also decided to use Java, one of the widely used programming languages in web development contexts.

Krüger et al. [9] conducted an empirical study with 60 open-source developers from 10 GitHub projects. They use an online survey methodology in this study and investigate the forgetting curve concept in the software engineering concept by focusing on repetition, the ratio of own code, change tracking, the average memory strength of the developer regarding the source code, and Ebbinghaus' forgetting curve applicability for software developers. As a result of the work, they found the Ebbinghaus forgetting curve applicable for software development only if time has to be considered. They also asserted that there is a comparatively strong relationship between repetitions and the familiarity, and when it is compared to the elapsed time, it can be more significant. Their work closely aligns the objectives of our study with a difference in terms of methodology. Since the research methodology applied in this study was a survey, the authors were only able to analyze the perceptions of forgetting among the participants and not able to measure the real effects of forgetting. On the other hand, in our study since we apply both a controlled experiment approach and a survey, we will be able to measure the effects of forgetting and then also compare these results with the participants' perceptions of forgetting.